\documentclass[conference]{IEEEtran}
\IEEEoverridecommandlockouts
\usepackage{cite}
\usepackage{amsmath,amssymb,amsfonts}
\usepackage{xcolor}
\usepackage{enumerate}
\usepackage{hyperref}

\newtheorem{defin}{Definition}[section]
\newtheorem{teo}{Theorem}[section]
\newtheorem{lema}{Lemma}[section]
\newtheorem{remark}{Remark}[section]
\newtheorem{cor}{Corollary}[section]

\newtheorem{fact}{Fact}

\def\K{\mathcal{K}}
\def\Ki{\K_{\infty}}
\def\KL{\mathcal{KL}}
\def\R{\mathbb{R}}
\def\N{\mathbb{N}}
\def\S{\mathcal{S}}
\def\T{\mathcal{T}}
\def\U{\mathcal{U}}

\def\mer{\hfill $\circ$}

\def\comp{{\scriptstyle\,\circ}\,}
\def\A{\mathcal{A}}
\def\AL{\A\mathcal{L}}
\def\I{\mathcal{I}}

\begin{document}

\title{A characterization of strong iISS\\ for time-varying impulsive systems\thanks{E-mails: \texttt{\{haimovich,cardone\}@cifasis-conicet.gov.ar}, \texttt{jmancill@itba.edu.ar}}}

\author{\IEEEauthorblockN{Hernan Haimovich}
\IEEEauthorblockA{\textit{Centro de Ciencias de la}\\ \textit{Informaci\'on y Sistemas (CIFASIS)} \\
\textit{CONICET -- UNR}\\
2000 Rosario, Argentina.}
\and
\IEEEauthorblockN{Jos\'e Luis Mancilla-Aguilar}
\IEEEauthorblockA{\textit{Centro de Sistemas y Control} \\
\textit{Instituto Tecnol\'ogico de Buenos Aires}\\
Av. Eduardo Madero 399,\\ Buenos Aires, Argentina.}
\and
\IEEEauthorblockN{Paula Cardone}
\IEEEauthorblockA{\textit{CIFASIS} \\
\textit{CONICET -- UNR}\\
Ocampo y Esmeralda,\\ 2000 Rosario, Argentina.}
}

\maketitle

\begin{abstract}
  For general time-varying or switched (nonlinear) systems, converse Lyapunov theorems for stability are not available. In these cases, the integral input-to-state stability (iISS) property is not equivalent to the existence of an iISS-Lyapunov function but can still be characterized as the combination of global uniform asymptotic stability under zero input (0-GUAS) and uniformly bounded energy input-bounded state (UBEBS). For impulsive systems, asymptotic stability can be weak (when the asymptotic decay depends only on elapsed time) or strong (when such a decay depends also on the number of impulses that occurred). This paper shows that the mentioned characterization of iISS remains valid for time-varying impulsive systems, provided that stability is understood in the strong sense.
\end{abstract}


\section{Introduction}

Impulsive systems are dynamical systems whose state evolves continuously most of the time but may exhibit jumps (discontinuities) at isolated time instants (see \cite{lakbai_book89}). The continuous evolution of the state (i.e. between jumps) is governed by ordinary differential equations. The time instants when jumps occur are part of the impulsive system definition and the after-jump value of the state vector is governed by a static (i.e. not differential) equation. Uniform asymptotic stability of the origin requires that the norm of the state decays asymptotically to zero as elapsed time advances. For an impulsive system, uniform asymptotic stability can be defined in two different ways, depending on whether the decay depends only on elapsed time (we use the name weak in this case) \cite{heslib_auto08} or also on the number of impulses that have occurred (strong) \cite{libnes_tac14}.

Input-to-state stability (ISS) \cite{sontag_tac89} and integral-ISS (iISS) \cite{sontag_scl98} are arguably the most important and useful state-space based nonlinear notions of stability for systems with inputs. The iISS property gives a state bound that is the sum of a decaying-to-zero term whose amplitude depends only on the initial state, and a term depending (nonlinearly) only on an integral of a nonlinear function of the input. The latter term can be interpreted as an input energy bound. As is the case with uniform asymptotic stability, for impulsive systems the decaying-to-zero term can take impulse occurrence into account or not, giving rise to two different ways of defining iISS (strong or weak). The weak iISS property is the most usual for impulsive systems \cite{heslib_auto08}, whereas the strong version is in agreement with iISS for hybrid systems \cite{norkha_mcss17}.

Several different sufficient conditions for weak iISS of impulsive systems involving time-invariant or time-varying flow and jump equations, with or without time delays, exist \cite{heslib_auto08,chezhe_auto09,liuliu_auto11,lizha_auto17,peng_ietcta18,liuhil_amc18,ninhe_is18,lili_mcs19}. However, to the best of our knowledge, conditions that are both necessary and sufficient only exist for strong iISS when the impulsive system can be posed as a time-invariant hybrid system where the (time-invariant) flow map, in addition, satisfies a convexity property with respect to the input variable \cite{norkha_mcss17}. For time-invariant nonimpulsive systems, iISS was shown to be equivalent to the combination of global uniform asymptotic stability under zero input (0-GUAS) and uniformly bounded-energy input bounded state (UBEBS). This characterization of iISS was extended to time-varying and switched (nonimpulsive) systems \cite{haiman_tac18}, and has been recently shown to remain valid for impulsive systems provided stability is understood in the weak sense and the number of jumps that occur in any given time interval is bounded in relation to the interval's length but irrespective of initial time \cite{haiman_aadeca18}.

In this paper, we show that the previously derived characterization of iISS (namely, iISS = 0-GUAS + UBEBS) remains valid for impulsive systems provided stability is understood in the strong sense and without having to bound the number of jumps as in the weak case. As was the case with the previous results \cite{haiman_aadeca18}, the current results apply to cases where both the ordinary differential equation defining continuous state evolution (i.e. the flow equation) and the static equation defining after-jump values (i.e. the jump equation) can be time-varying and lack time continuity. The results of \cite{haiman_aadeca18} are then shown to be a particular case of the current ones. 


\textbf{Notation.} $\N$, $\R$, $\R_{>0}$ and $\R_{\ge 0}$ denote the natural numbers, reals, positive reals and nonnegative reals, respectively. $|x|$ denotes the Euclidean norm of $x \in \R^p$. 
 We write $\alpha\in\K$ if $\alpha:\R_{\ge 0} \to \R_{\ge 0}$ is continuous, strictly increasing and $\alpha(0)=0$, and $\alpha\in\Ki$ if, in addition, $\alpha$ is unbounded. We write $\beta\in\KL$ if $\beta:\R_{\ge 0}\times \R_{\ge 0}\to \R_{\ge 0}$, $\beta(\cdot,t)\in\Ki$ for any $t\ge 0$ and, for any fixed $r\ge 0$, $\beta(r,t)$ monotonically decreases to zero as $t\to \infty$. For every $n\in\N$ and $r\ge 0$, we define the closed ball $B_r^n := \{x\in\R^n : |x| \le r\}$. A function $h : D \subset \R \times \R^n \to \R^n$ is said to be a Carath\'eodory function if $h(t,\xi)$ is measurable in $t$ for fixed $\xi$, continuous in $\xi$ for fixed $t$, and for every compact set $K \subset D$, there exists an integrable function $m_K(t)$ such that $|h(t,\xi)| \le m_K(t)$ for all $(t,\xi)\in K$ (see \cite[Sec.~I.5]{hale_book80}).

\section{Problem Statement}

\subsection{Impulsive systems}
\label{sec:prel-defs}

Consider the time-varying impulsive system with inputs
\begin{subequations}
  \label{eq:is}
  \begin{align}
    \label{eq:is-ct}
    \dot{x}(t) &=f(t,x(t),u(t)),\phantom{x(t^-)+g^-}\quad\text{for } t\notin \sigma,    \displaybreak[0] \\
    \label{eq:is-st}
    x(t) &=x(t^-) + g(t,x(t^-),u(t)),\phantom{f} \quad\text{for } t\in \sigma,
  \end{align}
\end{subequations}
where $t_0\ge 0$ is the initial time, $\sigma=\{\tau_k\}_{k=1}^{N}$, with $N$ finite or $N=\infty$, is a strictly increasing sequence of impulse times in $\R_{>0}$, the state variable $x(t)\in \R^n$, the continuous-time input variable $u(t)\in \R^m$ and $f$ (the flow map) and $g$ (the jump map) are functions from $\R_{\ge 0}\times \R^n\times \R^m$ to $\R^n$. The ordinary differential equation (\ref{eq:is-ct}) defines the continuous evolution of the state vector $x$ and (\ref{eq:is-st}) defines the value of $x$ at the impulse times. To ensure that the jumps in $x$ caused by (\ref{eq:is-st}) cannot occur infinitely frequently, it is assumed that $\tau_k\to \infty$ when $N=\infty$. By convention we define $\tau_0=0$ (however, $\tau_0$ is not considered an impulse time) and, when $N$ is finite, we set $\tau_{N+1}:=\infty$. We will employ $\I$ to denote the set of all these admissible impulse time sequences, i.e. $\I$ denotes the set of all strictly increasing sequences of positive real numbers that either have a finite number of elements or are unbounded. Let $\mathcal{U}$ be the set of all the functions $u:\R_{\ge 0}\to \R^m$ that are Lebesgue measurable and locally bounded. We will use the term ``input'' to refer to a pair $w=(u,\sigma) \in \U \times \I$ consisting of a continuous-time input $u$ and an admissible impulse-time sequence $\sigma$.  We assume that for each $u\in \U$ the map $f_u(t,\xi):=f(t,\xi,u(t))$ is a Carath\'eodory function and hence the (local) existence of solutions of the differential equation $\dot x(t)=f(t,x(t),u(t))$ is ensured (see \cite[Thm.~I.5.1]{hale_book80}).


A solution to (\ref{eq:is}) corresponding to an initial time $t_0$, an initial state $x_0\in \R^n$ and an input $w=(u,\sigma) \in \U \times \I$ is a right-continuous function $x:[t_0,T_x)\to \R^n$ such that:
\begin{enumerate}[i)]
\item $x(t_0)=x_0$; 
\item $x$ is a Carath\'eodory solution of the differential equation $\dot{x}(t)=f(t,x(t),u(t))$ on $[\tau_{k},\tau_{k+1})\cap [t_0,T_x)$ for all $0\le k \le N$; and 
\item for all $t\in \sigma \cap (t_0,T_x)$ it happens that $x(t)=x(t^-)+ g(t,x(t^-),u(t))$, where $x(t^-) := \lim_{s\to t^-} x(s)$.
\end{enumerate}
The solution $x$ is said to be maximally defined if no other solution $y$ satisfies $y(t) = x(t)$ for all $t\in [t_0,T_x)$ and has $T_y > T_x$. A solution $x$ is forward complete if $T_x=\infty$. We will use $\T(t_0,x_0,w)$ to denote the set of maximally defined solutions of (\ref{eq:is}) corresponding to initial time $t_0$, initial state $x_0$, and input $w$. We say that (\ref{eq:is}) is forward complete for a given $\sigma\in\I$ if for every $t_0 \ge 0$, $x_0 \in \R^n$ and $w=(u,\sigma)$ with $u\in\U$, any solution $x\in\T(t_0,x_0,w)$ is forward complete. Given $\sigma \in \I$, we define $n^\sigma_{(t_0,t]}$ to be the number of elements of $\sigma$ (i.e. the number of jumps) that lie in the interval $(t_0,t]$:
\begin{align}
  n^\sigma_{(t_0,t]} &:= \# \big[ \sigma\cap (t_0,t] \big].
\end{align}

\subsection{Stability definitions}
\label{sec:stab-defs}

Stability notions for systems with inputs that are uniform with respect to initial time, such as uniform ISS and iISS, bound the state trajectory in relation to initial state, elapsed time and input. In the context of impulsive systems, the input can be interpreted as having both a continuous-time and an impulsive component. 
Given an input $w = (u,\sigma)$ and $\rho_1,\rho_2\in\Ki$, we thus define
\begin{align}
  \label{eq:iiss-norm}
  \| w \|_{(\rho_1,\rho_2)} &:= \int_0^\infty \rho_1(|u(s)|) ds + \sum_{t\in\sigma} \rho_2(|u(t)|).
\end{align}
The quantity defined in (\ref{eq:iiss-norm}) can be loosely interpreted as a measure of the energy content of an input that has some impulsive behaviour at the time instants $t\in\sigma$.

We are interested in determining whether some stability property holds not just for a single impulse-time sequence $\sigma\in\I$ but also for some family $\S \subset \I$. 
We thus consider the uniform stability notions given in Definition~\ref{def:stab}. To simplify notation, for every interval $J \subset [0,\infty)$ and $u\in\U$, we define $u_J$ via $u_J(t) := u(t)$ if $t\in J$ and $u_J(t) := 0$ otherwise; for an input $w=(u,\sigma)$, we define $w_J := (u_J,\sigma)$. 
\begin{defin}
  \label{def:stab}
  Given $\S\subset\I$, we say that the impulsive system (\ref{eq:is})
  is
  \begin{enumerate}[a)]
  \item strongly 0-GUAS uniformly over (the family of impulse-time sequences) $\S$ if there exists $\beta \in \KL$ such that
    \begin{align}
      \label{eq:0-guas}
      |x(t)| &\le \beta(|x(t_0)|,t-t_0+n^\sigma_{(t_0,t]}) \quad \forall t\ge t_0,
    \end{align}
    for every $x\in\T(t_0,x_0,w_0)$ with $t_0 \ge 0$, $x_0 \in \R^n$
    and $w_0=(0,\sigma)$ with $\sigma\in\S$.
  \item UBEBS uniformly over $\S$ if there exist $\alpha,\rho_1,\rho_2\in\Ki$ and $c\ge 0$ such that
    \begin{align}
      \label{eq:cubebs}
      \alpha(|x(t)|) &\le |x(t_0)| + \| w_{(t_0,t]} \|_{(\rho_1,\rho_2)} + c \quad \forall t\ge t_0,
    \end{align}
    for every $x\in\T(t_0,x_0,w)$ with $t_0 \ge 0$, $x_0 \in \R^n$ and $w \in \U \times \S$. The pair $(\rho_1,\rho_2)$ will be referred to as an UBEBS gain.
  \item strongly iISS uniformly over $\S$ if there exist $\beta \in \KL$ and $\alpha,\rho_1,\rho_2 \in \Ki$ such that
    \begin{multline}
      \label{eq:ciiss}
      \alpha(|x(t)|) \le \beta(|x(t_0)|,t-t_0+n^\sigma_{(t_0,t]})\\
      + \| w_{(t_0,t]} \|_{(\rho_1,\rho_2)}
    \end{multline}
    for all $t\ge t_0$, for every $x\in\T(t_0,x_0,w)$ with $t_0 \ge 0$, $x_0 \in \R^n$ and $w \in \U \times \S$. The pair $(\rho_1,\rho_2)$ will be referred to as an iISS gain.
  \end{enumerate}
\end{defin}
\begin{remark}
Due to the blanket assumption we have made on $f$, any of the conditions (\ref{eq:0-guas}), (\ref{eq:cubebs}) or (\ref{eq:ciiss}) implies that the solution $x$ is forward complete. Suppose that $x$ is a solution satisfying (\ref{eq:0-guas}), (\ref{eq:cubebs}) or (\ref{eq:ciiss}) and that its maximal interval of definition is $[t_0,T)$ or $[t_0,T]$ with $T<\infty$. If $x(T)$ is defined, then the initial value problem $\dot{z}(t)=f(t,z(t),u(t))$, $z(T)=x(T)$ has a solution $z$ which is defined on some interval $[T,T+\delta)$ with $\delta>0$. In consequence, $x$ admits a prolongation defined on some interval $[t_0,T+\delta')$ with $\delta'>0$ small enough, which is absurd. If $x(T)$ is not defined, then, due to standard results on ordinary differential equations, $|x(t)|\to \infty$ as $t\to T^-$, but this is impossible since $x$ is bounded on $[t_0,T)$. \mer  
\end{remark}

The weak versions of 0-GUAS or iISS are obtained by replacing the second argument of the function $\beta$ in (\ref{eq:0-guas}) or (\ref{eq:ciiss}) by just $t-t_0$ (i.e. the number of jumps $n^\sigma_{(t_0,t]}$ does not appear).
If (\ref{eq:is}) is (weakly or strongly) 0-GUAS uniformly over $\S$, then under $u\equiv 0$ the state converges asymptotically to the origin. In the weak case, the convergence warranty depends on the elapsed time $t-t_0$ but is insensitive to the occurrence of jumps. In addition, this convergence is uniform over initial times and over impulse time sequences within the family $\S$. The uniform-over-$\S$ UBEBS property just imposes a bound on the state trajectory without necessarily guaranteeing convergence. The bound is uniform over initial times and over all $\sigma\in\S$, and depends on the initial state norm and the input energy. The uniform-over-$\S$ (weak or strong) iISS property imposes a bound that is also uniform over initial times and over all $\sigma\in\S$. This bound is formed by a term similar to the 0-GUAS property and another term equal to the input energy.



\section{Characterization of Strong iISS}
\label{sec:main-res}

\subsection{Main result}
\label{sec:ass}

We require the following definitions, as employed in \cite{haiman_aadeca18}.
\begin{defin}
  \label{def:A}
  A function $h : \R_{\ge 0} \times \R^n \times \R^m \to \R^n$ is said to belong to class $\AL$, written $h\in\AL$, if the following items hold:
  \begin{enumerate}[i)]
  \item \label{item:bound} there exist $\nu_h \in \K$ and a nondecreasing function $N_h:\R_{\ge 0}\to \R_{>0}$ such that $|h(t,\xi,\mu)|\le N_h(|\xi|)(1+\nu_h(|\mu|))$ for all $t\ge 0$, all $\xi \in \R^n$ and all $\mu\in \R^m$;
  \item for every $r>0$ and $\varepsilon >0$ there exists $\delta>0$ such that for all $t\ge 0$, $|h(t,\xi,\mu) - h(t,\xi,0)|<\varepsilon$ if $|\xi|\le r$ and $|\mu| \le \delta$.\label{item:fcont}
  \item  \label{item:lips} $h(t,\xi,0)$ is locally Lipschitz in $\xi$,     uniformly in $t$, i.e. for every $\xi\in\R^n$ there are an open ball $B$ containing $\xi$ and a constant $L\ge 0$ so that for every $\xi_1,\xi_2 \in B$ and $t\ge 0$ it happens that $|h(t,\xi_1,0) - h(t,\xi_2,0)| \le L |\xi_1-\xi_2|$.
  \end{enumerate}
\end{defin}
%

Our main result is the following.
\begin{teo}
  \label{UBEBSand0GASiffiiss}
  Consider the impulsive system (\ref{eq:is}), suppose that $f,g\in\AL$ and let $\S \subset \I$. Then, (\ref{eq:is}) is strongly iISS uniformly over $\S$ if and only if it is strongly 0-GUAS and UBEBS, both uniformly over $\S$.
\end{teo}
The proof of Theorem~\ref{UBEBSand0GASiffiiss} will be developed along Sections~\ref{sec:inter-res} and~\ref{sec:eps-del}.

\subsection{Intermediate results}
\label{sec:inter-res}

The proof of Theorem~\ref{UBEBSand0GASiffiiss} follows the same steps as that of the proof of Theorem~3.2 of \cite{haiman_aadeca18} but suitably modified for the strong case. For the sake of conciseness and to clarify the current contribution, we will emphasize the main differences and remove the parts that are identical or very similar.

The integral expression for the solution of (\ref{eq:is}) is given by:
\begin{multline}
  \label{eq:solintform}
  x(t) = x(t_0) + \int_{t_0}^t f(s,x(s),u(s)) ds +\displaybreak[0]\\
  +\sum_{\tau\in\sigma\cap(t_0,t]} g(\tau,x(\tau^-),u(\tau)).
\end{multline}
The proof of our main result requires the generalization of Gronwall inequality for continuous functions with isolated jumps given as Lemma~3.1 in \cite{haiman_aadeca18}. We copy the corresponding statement here for simplicity.
\begin{lema}[{\cite[Lemma~3.1]{haiman_aadeca18}}]
  \label{lem:ggi}
  Let $0 \le t_0 < T \le \infty$ and let $y:[t_0,T) \to \R$ be a right-continuous function having a finite left-limit at every discontinuity instant. Suppose that the points of discontinuity of $y$ can be arranged into a sequence $\sigma\in\I$. Let $p\in\R$ and $q_1,q_2 \ge 0$. If $y$ satisfies
  \begin{align}
    y(t) &\le p + q_1 \int_{t_0}^t y(s) ds + q_2 \sum_{s\in\sigma\cap (t_0,t]} y(s^-)
  \end{align}
  for all $t \in [t_0,T)$, then in the same time interval $y$ also satisfies
  \begin{align}
    y(t) &\le p (1+q_2)^{n^\sigma_{(t_0,t]}}\cdot e^{q_1 (t-t_0)}.
  \end{align}
\end{lema}

We will also require Lemma~3.2 of \cite{haiman_aadeca18} (which is a generalization of Lemma~3 in \cite{haiman_tac18}) suitably modified for the strong 0-GUAS case. The proof is a very minor modification of the corresponding proof in \cite{haiman_aadeca18} and hence omitted.
\begin{lema}[cf.~Lemma~3.2 in \cite{haiman_aadeca18}]
  \label{lem:genlem3}
  Let $\S\subset\I$, let the impulsive system (\ref{eq:is}) be strongly 0-GUAS uniformly over $\S$ and let $\beta\in\KL$ characterize the strong 0-GUAS property. Suppose that $f,g \in \AL$ and let $\nu_f$ and $\nu_g$ be, respectively, the functions corresponding to $f$ and $g$ as per item~\ref{item:bound}) of Definition~\ref{def:A}. Let $\chi_f,\chi_g \in \Ki$ satisfy $\chi_f \ge \nu_f$ and $\chi_g \ge \nu_g$. Then, for every $r>0$ and every $\eta>0$, there exist $L=L(r)$ and $\kappa = \kappa(r,\eta)$ such that if $x\in\T(t_0,x_0,w)$ with $t_0 \ge 0$, $x_0\in\R^n$, $w=(u,\sigma) \in \U\times\S$ satisfies $|x(t)| \le r$ for all $t\ge t_0$, then also
  \begin{multline}
    \label{eq:genlem3bnd}
    |x(t)| \le \beta(|x_0|,t-t_0+n^\sigma_{(t_0,t]}) +
    \left[{\scriptstyle (t-t_0+n^\sigma_{(t_0,t]})}\eta \right. \\ \left.+ \kappa \|w_{(t_0,t]}\|_{(\chi_f,\chi_g)} \right]  \scriptstyle (1+L)^{n^\sigma_{(t_0,t]}}\cdot e^{L(t-t_0)}. 
  \end{multline}
\end{lema}
The only difference with respect to the corresponding bound in Lemma~3.2 of \cite{haiman_aadeca18} is the inclusion of the number of jumps $n^\sigma_{(t_0,t]}$ within the second argument of $\beta$ in (\ref{eq:genlem3bnd}). The corresponding proof is almost identical. 

The proof of our main result also requires a suitably modified version of Lemma~3.3 of \cite{haiman_aadeca18}. In this case, the removal of the assumption on the boundedness of the number of jumps in a given interval, given by the uniform incremental boundedness (UIB) property in \cite{haiman_aadeca18}, makes the corresponding proof sufficiently different so as to include it here. 
\begin{lema}
  \label{lem:0UBEBS} 
  Consider the impulsive system (\ref{eq:is}), suppose that $f,g\in\AL$ and let $\S\subset\I$. If (\ref{eq:is}) is strongly 0-GUAS and UBEBS, both uniformly over $\S$, then there exist $\tilde\alpha,\tilde\rho_1,\tilde\rho_2\in \Ki$ for which the estimate (\ref{eq:0UBEBS}) holds for every $x \in \T(t_0,x_0,w)$ with $t_0\ge 0$, $x_0\in \R^n$ and $w\in \U \times \S$.
  \begin{align} 
    \label{eq:0UBEBS}
    \tilde\alpha(|x(t)|) \le |x(t_0)| + \|w_{(t_0,t]}\|_{(\tilde\rho_1,\tilde\rho_2)} \quad \forall t\ge t_0.
 \end{align}
\end{lema}
\begin{IEEEproof}
  Let $\alpha$, $\rho_1$, $\rho_2$ and $c$ be as in the estimate~(\ref{eq:cubebs}). Let $\tilde\rho_1 := \max\{\rho_1,\nu_f\}$ and $\tilde\rho_2 := \max\{\rho_2,\nu_g\}$. For $r\ge 0$ define
  \begin{align*}
    \bar\alpha(r) &:= \sup_{x\in \mathcal{T}(t_0,x_0,w),\; t\ge t_0\ge 0,\; |x_0|\le r,\;  w\in \U\times\S, \; \|w\|\le r} |x(t)|
  \end{align*}
where $\|w\|:=\|w\|_{(\tilde\rho_1,\tilde\rho_2)}$. From this definition, it follows that $\bar\alpha$ is nondecreasing and from (\ref{eq:cubebs}) that it is finite for all $r\ge 0$. 
Let $\beta\in \KL$ be the function which characterizes the uniform-over-$\S$ strong 0-GUAS property of (\ref{eq:is}). From the latter property, it follows that $\bar\alpha(0) = 0$. Next, we show that $\lim_{r\to 0^+}\bar\alpha(r)=0$. Let $r^*=\alpha^{-1}(2+c)$ and $L=L(r^*)>0$ be given by Lemma~\ref{lem:genlem3}. Let $\varepsilon>0$ be arbitrary. Pick $0<\delta_1<1$ such that $\delta_1 \le \beta(\delta_1,0)<\varepsilon/2$ and $\tilde{T}>0$ such that $\beta(\delta_1,\tilde{T})<\delta_1/2$. Define $\eta = \frac{\delta_1}{4 (\tilde{T}+1)} e^{-\tilde{L}(\tilde{T}+1)}$, with $\tilde{L} = \max \left\{L, \log(1+L) \right\}$ and let $\kappa=\kappa(r^*,\eta)>0$ be given by Lemma~\ref{lem:genlem3}. Last, pick $0<\delta_2<1$ such that
  $\delta_2 < \frac{\delta_1}{4 \kappa (\tilde{T} + 1)} e^{-\tilde{L}(\tilde{T} + 1)}$.
For every $j\in\N_0$, define 
\begin{align*}
  t_{j+1} := \inf \left\{ t>t_{j} : t-t_{j}+n^\sigma_{(t_j,t]} \geq \tilde{T} \right\}
\end{align*}
and consider the intervals $I_j = [t_j,t_{j+1})$. Note that $t_{j+1} > t_j$ for every $j\in\N_0$. 
By definition of $t_{j+1}$ and since $t\mapsto n_{(t_j,t]}^\sigma$ is right-continuous, it follows that
for all $j \in \N_0$,
\begin{align*}
  \tilde{T} \leq t_{j+1}-t_{j}+n^\sigma_{(t_j,t_{j+1}]} &\leq \tilde{T}+1,\quad\text{and}\\
  t-t_{j}+n^\sigma_{(t_j,t]} &\leq \tilde{T}+1 \quad\forall  t \in I_j.
\end{align*}
We claim that 
$\lim_{j\to \infty}t_{j} = \infty$.  For a contradiction, suppose that $\lim_{j\to \infty}t_{j} = M < \infty$. As every convergent sequence is a Cauchy sequence, for every $\rho>0$ there exists $N = N(\rho) \in \N$ such that $|t_{j+1}-t_j|<\rho$ for all $j \ge N$. But $t_{j+1}-t_j+n^\sigma_{(t_j,t_{j+1}]} \geq \tilde{T}>0$ and hence $0 < \tilde{T} \leq \rho+n^\sigma_{(t_j,t_{j+1}]}$ and $\tilde{T}-\rho<n^\sigma_{(t_j,t_{j+1}]}$. Taking $\rho<\tilde{T}$ we have that $0<\tilde{T}-\rho < n^\sigma_{(t_j,t_{j+1}]}$ and thus $n^\sigma_{(t_j,t_{j+1}]} \geq 1$ for all $j\ge N$. Then, $n^\sigma_{(t_N,M)} = \sum_{j=N}^{\infty} n^\sigma_{(t_j,t_{j+1}]} \ge \sum_{j=N}^{\infty} 1 = \infty$, contradicting the assumption that $\sigma$ has no finite accumulation points. Therefore, $\lim_{j\to \infty}t_{j} = \infty$. 

For every $x\in \T(t_0,x_0,w)$, with $t_0\ge 0$, $|x_0|\le \delta_1$, $w \in \U\times\S$ and $\|w\| \le \delta_2$, we also have $x \in \T(t_j,x(t_j),w)$ for all $j\in\N$. By induction, we will show that $|x(t)| \le \varepsilon$ for all $t\in I_j = [t_j,t_{j+1})$ and that $|x(t_{j+1})| < \delta_1$. For $j=0$ and applying Lemma~\ref{lem:genlem3}, it follows that for all $t \in I_{0}$, we have
\begin{align*}
  |x(t)| &\le \beta(|x_{0}|,t-t_{0}+n^\sigma_{(t_0,t]})\\
  &\hspace{10mm}+[(t-t_{0}+n^\sigma_{(t_0,t]})\eta + \kappa\|w\|] e^{\tilde{L}(t-t_{0}+n^\sigma_{(t_0,t]})}\\
  &\le \beta(\delta_{1},0) + [(\tilde{T}+1)\eta + \kappa\delta_2] e^{\tilde{L}(\tilde{T}+1)} < \frac{\varepsilon}{2} + \frac{\delta_1}{2} < \varepsilon,
\end{align*}
and that
\begin{align*}
  |x(t_1)| &\le \beta(|x_{0}|, t_{1}-t_{0}+n^\sigma_{(t_0,t_{1}]})\\ &\hspace{5mm}+[(t_{1}-t_{0}+n^\sigma_{(t_0,t_{1}]})\eta + \kappa\|w\|] e^{\tilde{L}(t_{1}-t_{0}+n^\sigma_{(t_0,t_{1}]})}\\
  &\le \beta(\delta_{1},\tilde{T})+[(\tilde{T}+1)\eta + \kappa\delta_2] e^{\tilde{L}(\tilde{T}+1)} < \delta_{1}.
\end{align*}
So our induction assumption holds for $j=0$. Next, suppose that it holds for arbitrary $j\in N_0$. Applying Lemma~\ref{lem:genlem3}, then for $t\in I_{j+1}$ we have that
\begin{align*}
  |x(t)| &\le \beta(|x(t_{j+1})|,0) + [(\tilde{T}+1)\eta + \kappa\delta_2] e^{\tilde{L}(\tilde{T}+1)}\\
  &\le \beta(\delta_{1},0) + [(\tilde{T}+1)\eta + \kappa\delta_2] e^{\tilde{L}(\tilde{T}+1)} < \frac{\varepsilon}{2} + \frac{\delta_1}{2} < \varepsilon,
\end{align*}
where we have used the fact that $|x(t_{j+1})| \le \delta_1$, and that
\begin{align*}
  |x(t_{j+2})| &\le \beta(|x(t_{j+1})|, t_{j+2}-t_{j+1}+n^\sigma_{(t_{j+1},t_{j+2}]})\displaybreak[0]\\ &\hspace{5mm}+[(\tilde{T}+1)\eta + \kappa\delta_2] e^{\tilde{L}(\tilde{T}+1)}\\
  &\le \beta(\delta_{1},\tilde{T})+[(\tilde{T}+1)\eta + \kappa\delta_2] e^{\tilde{L}(\tilde{T}+1)} < \delta_{1}.
\end{align*}
Hence our induction assumption holds for $j+1$. As a consequence, $|x(t)| \le \varepsilon$ must hold for all $t \ge t_0$.
%
%
Thus, if $\delta=\min\{\delta_1,\delta_2\}$, for all
$x\in \T(t_0,x_0,w)$, with $t_0\ge 0$, $|x_0|\le \delta$, $w\in \U\times \S$ with
$\|w\|\le \delta$, we have $|x(t)|\le \varepsilon$ for all $t\ge t_0$. Therefore,
$\bar\alpha(r)\le \bar\alpha(\delta)<\varepsilon$ for all
$0<r<\delta$ and $\lim_{r\to 0^+}\bar\alpha(r)=0$.
 
Since $\bar\alpha$ is nondecreasing and
$\lim_{r\to 0^+}\bar\alpha(r)=0$ there exists $\hat \alpha \in \Ki$ such
that $\hat \alpha(r)\ge \bar\alpha(r)$ for all $r\ge 0$. Let
$x \in \T(t_0,x_0,w)$ with $t_0\ge 0$, $x_0 \in \R^n$ and $w\in \U \times \S$. Let $t\ge t_0$. 
Due to causality, there exists
$x^*\in \T(t_0,x_0,w_{(t_0,t]})$ such that $x^*(\tau)=x(\tau)$ for all
$\tau \in [t_0,t]$. By using the definition of $\bar\alpha$ and the
fact that $\hat \alpha(r)\ge \bar\alpha(r)$, we then have
  $|x(t)| = |x^*(t)| \le \hat\alpha(|x_0|) + \hat\alpha(\|w_{(t_0,t]}\|)$.
Define $\tilde\alpha \in \Ki$ via $\tilde\alpha(s) = \hat\alpha^{-1}(s)/2$. Applying $\tilde\alpha$ to both sides of the preceding inequality and using the fact that $\tilde\alpha(a+b) \le \tilde\alpha(2a) +\tilde\alpha(2b)$, we reach
  $\tilde\alpha(|x(t)|) \le |x_0| + \|w_{(t_0,t]}\|$,
which establishes the result.
\end{IEEEproof}


\subsection{Proof of Theorem~\ref{UBEBSand0GASiffiiss}}
\label{sec:eps-del}

The proof of our main result requires the following $\epsilon$-$\delta$ characterization of the uniform-over-$\S$ strong iISS property. The statement follows from suitable modification of that of Theorem~3.1 of \cite{haiman_aadeca18}. Whether this characterization holds or not under such a modification is a nontrivial question. We hence provide the proof in the Appendix.
\begin{teo}
  \label{thm:eps-delta}
  Let $\rho_1,\rho_2\in\Ki$ and $\S\subset\I$. Consider the notation $\|w\| = \|w\|_{(\rho_1,\rho_2)}$ and for $r \ge 0$, $B_r^\S := \{w\in\U\times\S : \| w \| \le r\}$. Then, system~(\ref{eq:is}) is strongly iISS uniformly over $\S$ with iISS gain $(\rho_1,\rho_2)$ if and only if 
  the following conditions hold:
  \begin{enumerate}[i)]
  \item For every $T\ge 0$, $r\ge 0$, $s\ge 0$, there exists $C>0$ such that every $x\in\T(t_0,x_0,w)$ with $t_0 \ge 0$, $x_0 \in B_r^n$ and $w \in B_s^\S$ satisfies $|x(t)| \le C$ for all $t \ge t_0$ such that $t + n^\sigma_{(t_0,t]} \le t_0+T$.\label{item:fc}
  \item For each $\epsilon > 0$, there exists $\delta > 0$ such that every $x\in\T(t_0,x_0,w)$ with $t_0 \ge 0$, $x_0 \in B_\delta^n$ and $w \in B_\delta^\S$ satisfies $|x(t)| \le \epsilon$ for all $t\ge t_0$.\label{item:sisicss}
  \item There exists $\tilde\alpha\in\Ki$ such that for every $r,\epsilon > 0$ there exists $T>0$ so that for every $x\in\T(t_0,x_0,w)$ with $t_0 \ge 0$, $x_0\in B_r^n$ and $w \in \U\times\S$, then\label{item:gatt}
      $\tilde\alpha(|x(t)|) \le \epsilon + \|w\|$
    for all $t\ge t_0$ such that $t + n^\sigma_{(t_0,t]} \ge t_0+T$.
  \end{enumerate}
\end{teo}

We may finally provide a proof to our main result.
\begin{IEEEproof}[Proof of Theorem~\ref{UBEBSand0GASiffiiss}]

($\Rightarrow$) Considering $w=(u,\gamma)$ with $u=0$, the estimate (\ref{eq:ciiss}) reduces to $\alpha(|x(t)|) \le \beta(|x(t_0)|,t-t_0+n^\sigma_{(t_0,t]})$ and hence $|x(t)| \le \alpha^{-1}(\beta(|x(t_0)|,t-t_0+n^\sigma_{(t_0,t]}))$. The function $\tilde\beta := \alpha^{-1} \comp \beta$ satisfies $\tilde\beta \in \KL$, and hence (\ref{eq:0-guas}) follows with $\beta$ replaced by $\tilde\beta$. Therefore, clearly strongly iISS implies strongly 0-GUAS, both uniformly over $\S$.

Consider $\beta \in \KL$ from (\ref{eq:ciiss}), define $\beta_0 \in \Ki$ via $\beta_0(r) = \beta(r,0)$. Then, $|x(t)| \le \alpha^{-1} \left[ \beta_{0}(|x(t_0)|) + \|w_{(t_0,t]}\|_{(\rho_1,\rho_2)} \right]$. 
Define $\psi\in\Ki$ via $\psi(r) = \min \left\{\beta_0^{-1}(\alpha(r)/2),  \alpha(r)/2  \right\}$. Applying $\psi$ to each side of the latter inequality and using the fact that $\phi(a+b) \le \phi(2a) + \phi(2b)$ for every $\phi\in\K$ and $a,b\ge 0$, yields
\begin{align*}
  \psi(|x(t)|) &\le \psi\comp\alpha^{-1}\left[ \beta_{0}(|x(t_0)|) + \|w_{(t_0,t]}\|_{(\rho_1,\rho_2)} \right]\\
  &\le \psi\comp\alpha^{-1} [ 2\beta_{0}(|x(t_0)|) ] + \psi\comp\alpha^{-1} [2\|w_{(t_0,t]}\|_{(\rho_1,\rho_2)} ]\\
  &\le |x(t_0)| + \|w_{(t_0,t]}\|_{(\rho_1,\rho_2)},
\end{align*}
and hence (\ref{eq:cubebs}) follows with $\alpha$ replaced by $\psi$. We have shown that strong iISS implies UBEBS, both uniformly over $\S$.

($\Leftarrow$)
Let $\tilde\alpha,\tilde\rho_1,\tilde\rho_2\in\Ki$ be given by Lemma~\ref{lem:0UBEBS}, so that (\ref{eq:0UBEBS}) is satisfied. We will prove that (\ref{eq:is}) is strongly iISS uniformly over $\S$ with iISS gain $(\tilde\rho_1,\tilde\rho_2)$ by establishing each of the items of Theorem~\ref{thm:eps-delta}.

\ref{item:fc}) Let $T\ge 0$, $r\ge 0$ and $s\ge 0$. Let $x \in \T(t_0,x_0,w)$ with $t_0\ge 0$, $x_0\in B_r^n$, $w\in B^\S_s$. From (\ref{eq:0UBEBS}) we have: $\tilde\alpha(|x(t)|) \le |x(t)| + \|w_{(t_0,t]}\|_{(\tilde\rho_1,\tilde\rho_2)} \le r+s$, and hence $|x(t)| \le \tilde\alpha^{-1}(r+s) =: C$ for all $t\ge t_0$. This establishes item~\ref{item:fc}) of Theorem~\ref{thm:eps-delta}.

\ref{item:sisicss}) Let $\epsilon > 0$. Let $\delta = \tilde\alpha(\epsilon)/2$. Then, if $x \in \T(t_0,x_0,w)$ with $t_0\ge 0$, $x_0\in B_\delta^n$ and $w\in B_\delta^\S$, from (\ref{eq:0UBEBS}) then $\tilde\alpha(|x(t)|) \le |x_0| + \|w_{(t_0,t]}\|_{(\tilde\rho_1,\tilde\rho_2)} \le \delta + \delta = 2 \delta$. It follows that $|x(t)| \le \tilde\alpha^{-1}(2\delta) = \epsilon$ for all $t\ge t_0$. This establishes item~\ref{item:sisicss}) of Theorem~\ref{thm:eps-delta}.

\ref{item:gatt}) Let $\alpha = \tilde\alpha/2 \in\Ki$. Let $r,\epsilon > 0$ and let $x \in \T(t_0,x_0,w)$ with $t_0\ge 0$, $x_0\in B_r^n$ and $w\in \U\times\S$. 
We distinguish two cases:
\begin{enumerate}[(a)]
\item $\|w\| \ge r$,
\item $\|w\| < r$.
\end{enumerate}
In case (a), from (\ref{eq:0UBEBS}) we have $\tilde\alpha(|x(t)|) \le |x_0| + \|w_{(t_0,t]}\| \le r + \|w\| \le 2\|w\|$, hence $\alpha(|x(t)|) = \frac{\tilde\alpha(|x(t)|)}{2} \le \|w\| \le \epsilon + \|w\|$ for all $t\ge t_0$.

Next, consider case (b). From (\ref{eq:0UBEBS}), we have  $\tilde\alpha(|x(t)|) \le r + \|w\| < 2r =: \tilde r$ for all $t\ge t_0$. Let $\beta \in \KL$ characterize uniform-over-$\S$ strong 0-GUAS property, so that (\ref{eq:0-guas}) is satisfied under zero input, and let $L = L(\tilde r) > 0$ be given by Lemma~\ref{lem:genlem3}. Define $\tilde L := \max\{L,\log(1+L)\}$, let $\tilde\epsilon = \epsilon$ and $\tilde T>0$ satisfy $\beta(\tilde r,\tilde T) < \tilde\epsilon /2$. Define $\eta=\frac{\tilde \epsilon}{4 (\tilde T+1)} e^{-\tilde L (\tilde T+1)}$. Let $\kappa = \kappa(\tilde r,\eta) > 0$ be given by Lemma~\ref{lem:genlem3}. Let $\delta = \frac{\tilde \epsilon}{4\kappa}e^{-\tilde L (\tilde T + 1)}$. Define $N := \left\lceil \frac{r}{\delta} \right\rceil$ and  $T:= N (\tilde T+1)$, where $\lceil s\rceil$ denotes the least integer not less than $s\in \R$.
Let $s_0 := t_0$ and for $i=1$ to $N$, define
\begin{align*}
  s_i := \inf \{ t \ge s_{i-1} : t - s_{i-1} + n^\sigma_{(s_{i-1},t]} \ge \tilde T\}.
\end{align*}
Then, for $i=1,\ldots,N$ we have $s_{i-1} < s_i < \infty$ and
\begin{align}
  \label{eq:Tnbnd}
  \tilde T \le s_i - s_{i-1} + n^\sigma_{(s_{i-1},s_i]} \le \tilde T + 1.
\end{align}
Consider the intervals $I_i=[s_{i-1}, s_{i}]$, with $i=1,\ldots, N$. We claim that there exists $j\le N-1$ for which $\|w_{(s_j,s_{j+1}]}\| \le \delta$. For a contradiction, suppose that $\|w_{(s_j,s_{j+1}]}\| > \delta$ for all $0 \le j \le N-1$. Then, $\|w\| \ge \|w_{(s_0,s_{N}]}\| = \sum_{j=0}^{N-1} \|w_{(s_j,s_{j+1}]}\| > N\delta \ge r$, contradicting case (b). Therefore, let $0\le j\le N-1$ be such that $\|w_{(s_j,s_{j+1}]}\| \le \delta$.

Since $x\in \T(s_{j},x(s_j),w)$ and $|x(t)|\le \tilde r$ for all $t\ge s_j$, from Lemma~\ref{lem:genlem3} and using the bounds (\ref{eq:Tnbnd}), it follows that
  \begin{multline*}
    |x(s_{j+1})| \le \beta(|x(s_j)|, \tilde T)+\\
    \left[ (\tilde T + 1) \eta + \kappa \|w_{(s_j,s_{j+1}]}\| \right] e^{\tilde L(\tilde T+1)} \\
    \le \beta(\tilde r,\tilde T) + [(\tilde T + 1) \eta + \kappa \delta] e^{\tilde L (\tilde T+1)} \le \tilde\epsilon.
  \end{multline*}
Therefore, using (\ref{eq:0UBEBS}) with $t_0$ replaced by $s_{j+1}$, we reach
\begin{align}
  \label{eq:xtbndfin}
  \tilde\alpha(|x(t)|) \le |x(s_{j+1})| + \| w_{(s_{j+1},t]} \| \le \tilde\epsilon + \|w\|
\end{align}
for all  $t\ge s_{j+1}$ and hence also for all $t\ge s_N$. Since $\sum_{i=1}^N s_i - s_{i-1} + n^\sigma_{(s_{i-1},s_i]} = s_N - s_0 + n^\sigma_{(s_0,s_N]} \le N(\tilde T+1) = T$, then $t + n^\sigma_{(t_0,t]} \ge t_0 + T$ implies that $t\ge s_N$. Therefore, (\ref{eq:xtbndfin}) holds for all $t\ge t_0$ for which $t + n^\sigma_{(t_0,t]} \ge t_0 + T$. Since $\alpha =\tilde\alpha/2 \le \tilde\alpha$, it follows that item~\ref{item:gatt}) of Theorem~\ref{thm:eps-delta} also is satisfied.
\end{IEEEproof}

\subsection{Previous results as a particular case}
\label{sec:prev-res-part-case}
In this section we will show that the main result in \cite{haiman_aadeca18}, namely Theorem 3.2 in \cite{haiman_aadeca18}, is a Corollary of Theorem \ref{UBEBSand0GASiffiiss}. We recall that a subset $\S\subset \mathcal{I}$ is uniformly incrementally bounded (UIB) if there exists a nondecreasing function $\phi : \R_{> 0} \to \R_{\ge 0}$ so that $n^\sigma_{(t_0,t]} \le \phi(t-t_0)$ for every $\sigma\in\S$ and all $t>t_0\ge 0$ (see Definition 3.2 in \cite{haiman_aadeca18}).

\begin{cor}(\cite[Thm.~3.2]{haiman_aadeca18}) \label{cor:aadeca} Consider the impulsive system (\ref{eq:is}) and suppose that $f,g\in\AL$. Let $\S \subset \I$ be a UIB set of impulse time sequences. Then, (\ref{eq:is}) is weakly iISS uniformly over $\S$ if and only if it is weakly 0-GUAS and UBEBS, both uniformly over $\S$.
\end{cor}
\begin{IEEEproof}
The proof of the only if part is straightforward and does not require the UIB hypothesis. As for the if part, assume that (\ref{eq:is}) is weakly 0-GUAS and UBEBS and that $\S$ is UIB. Let $\beta\in \KL$ be the function that characterizes the weak 0-GUAS stability property of the system (\ref{eq:is}). Let $\phi$ be the function appearing in the definition of the UIB property. Due to Lemma 6.1 in \cite{manhai_tac19arxiv}, there exists $\hat{\beta}\in \KL$ such that
\begin{align}
 \beta(r,s)\le \hat{\beta}(r,s+\phi(s)),\quad \forall (r,s)\in \R_{\ge 0}^2.
\end{align}
Then, for every $x\in \T(t_0,x_0,w_0)$ with $t_0 \ge 0$, $x_0 \in \R^n$ and $w_0=(0,\sigma)$ with $\sigma\in\S$ we have that for all $t\ge t_0$
\begin{align*}
      |x(t)| &\le \beta(|x(t_0)|,t-t_0) 
      \le \hat{\beta}(|x(t_0)|,t-t_0+\phi(t-t_0)) \\
      & \le \hat{\beta}(|x(t_0)|,t-t_0+n^\sigma_{(t_0,t]}).
    \end{align*}
So (\ref{eq:is}) is strongly 0-GUAS uniformly over $\S$. Applying Theorem~\ref{UBEBSand0GASiffiiss} it follows that (\ref{eq:is}) is then strongly iISS and therefore weakly iISS, both uniformly over $\S$.  
\end{IEEEproof}

\section{Conclusions}
\label{sec:concl}

We have addressed the characterization of the integral input-to-state stability property in terms of global uniform asymptotic stability under zero input and a uniformly bounded-energy input bounded state property. We have shown that this characterization remains valid for impulsive systems with time-varying flow and jump maps if both global uniform stability and integral input-to-state stability are understood in the strong sense. This characterization was established under a partial Lipschitz continuity assumption on the jump map [see item~\ref{item:lips}) of Definition~\ref{def:A}]. Future work is aimed at removing this assumption and establishing relationships between the ISS and iISS properties for impulsive systems.


\appendix

\subsection{Proof of Theorem~\ref{thm:eps-delta}}

  Necessity is straightforward, so we just establish sufficiency. 
Let $\tilde\alpha\in\Ki$ and $T>0$ be given by item~\ref{item:gatt}), the latter in correspondence with $r>0$ and $\epsilon = 1$. Let $C$ be given by item~\ref{item:fc}) in correspondence with $s=r$ and $T$. From items~\ref{item:fc}) and \ref{item:gatt}), we then have, whenever $t_0 \ge 0$, $x_0\in B_r^n$ and $w\in B_r^\S$, 
\begin{align*}
  |x(t)| \le C, &\quad\forall t\ge t_0,\ t + n^\sigma_{(t_0,t]} \le t_0+T,\\
  \tilde\alpha(|x(t)|) \le 1 + \|w\|, &\quad\forall t\ge t_0,\ t + n^\sigma_{(t_0,t]} > t_0 + T.
\end{align*}
It follows that $\tilde\alpha(|x(t)|) \le \tilde\alpha(C) + 1 + \|w\|$ for all $t\ge t_0$. 

Let $\phi(r) := \inf \{\tilde C\ge 0 : \tilde\alpha(|x(t)|) \le \tilde C,\ \forall x \in \T(t_0,x_0,w), \forall t \ge t_0 \ge 0, \forall x_0 \in B_r^n, \forall w\in B_r^\S \}$. By the previous analysis, then $\phi(r) \le \tilde\alpha(C) + 1 + r < \infty$ for all $r\ge 0$. Also, $\phi$ is nondecreasing and $\tilde\alpha(|x(t)|) \le \phi(|x(t_0)|) + \phi(\|w\|)$ for all $t\ge t_0$ whenever $x \in \T(t_0,x_0,w)$ with $t_0 \ge 0$, $x_0 \in \R^n$ and $w \in \U \times \S$. From item~\ref{item:sisicss}), it follows that $\lim_{r\searrow 0} \phi(r) = 0$. There thus exists $\eta\in\Ki$ such that $\phi\le\eta$ and then
\begin{align}
  \label{eq:talpleta}
  \tilde\alpha(|x(t)|) \le \eta(|x(t_0)|) + \eta(\|w\|) \quad\text{for all }t\ge t_0,
\end{align}
whenever $x \in \T(t_0,x_0,w)$ with $t_0 \ge 0$, $x_0 \in \R^n$ and $w \in \U \times \S$. Let $\psi,\alpha\in\Ki$ be defined via $\psi(s) = \eta^{-1}(s/2)$ and $\alpha = \min\{\tilde\alpha,\psi\comp\tilde\alpha\}$. Then, applying $\psi$ to (\ref{eq:talpleta}) and using the inequality $\psi(a+b) \le \psi(2a) + \psi(2b)$, it follows that
\begin{align}
  \label{eq:alple}
  \alpha(|x(t)|) \le |x(t_0)|+\|w\|\quad \text{for all }t\ge t_0,
\end{align}
whenever $x \in \T(t_0,x_0,w)$ with $t_0 \ge 0$, $x_0 \in \R^n$ and $w \in \U \times \S$.
Define
\begin{multline*}
  T_{r,\epsilon} := \inf\big\{ \tau\ge 0 : \alpha(|x(t)|) \le \epsilon + \|w\|,\\ \forall t\ge t_0, t + n^\sigma_{(t_0,t]} \ge t_0 + \tau,\\ \forall x\in\T(t_0,x_0,w), \forall t_0 \ge 0,  \forall x_0 \in B_r^n, \forall w \in \U \times \S \big\}.
\end{multline*}
By item~\ref{item:gatt}) and since $\alpha \le \tilde\alpha$, then $T_{r,\epsilon} < \infty$ for every $r,\epsilon > 0$. Moreover, $T_{r,\epsilon}$ is nondecreasing in $r$ for fixed $\epsilon>0$ and nonincreasing in $\epsilon$ for fixed $r>0$. By~(\ref{eq:alple}), then $T_{r,\epsilon} \to 0$ as $\epsilon \to \infty$ for fixed $r>0$.
\begin{fact}
  $T_{r,\epsilon}$ can be strictly upper bounded by $\bar T_{r,\epsilon}$ with the following properties:
  \begin{enumerate}[a)]
  \item For each fixed $r>0$, $\bar T_{r,\cdot} : \R_{>0} \to \R_{>0}$ is continuous, strictly decreasing, and onto, so that $\lim_{\epsilon\searrow 0} \bar T_{r,\epsilon} = \infty$ and $\lim_{\epsilon \to \infty} \bar T_{r,\epsilon} = 0$.
  \item For each fixed $\epsilon > 0$, $\bar T_{\cdot,\epsilon}$ is strictly increasing and $\lim_{r\to\infty} \bar T_{r,\epsilon} = \infty$.
  \end{enumerate}
\end{fact}
Let $\psi_r$ denote the inverse function of $\bar T_{r,\epsilon}$ considered as a function of $\epsilon$ for fixed $r> 0$. For every $r>0$, then $\psi_r$ is continuous on $\R_{>0}$ and $\lim_{s\searrow 0} \psi_r(s) = \infty$. By definition of $T_{r,\epsilon}$ and since $\bar T_{r,\epsilon} > T_{r,\epsilon}$, we have that
\begin{multline}
  \label{eq:impl1}
  t_0 \ge 0, x_0 \in B_r^n, w\in\U\times\S, x\in\T(t_0,x_0,w), t \ge t_0, \text{ and}\\
 t + n^\sigma_{(t_0,t]} \ge t_0 + \bar T_{r,\epsilon} \quad \Rightarrow \quad \alpha(|x(t)|) \le \epsilon + \|w\|
\end{multline}
Note that $t - t_0 + n^\sigma_{(t_0,t]} = \bar T_{r,\epsilon}$ is equivalent to $\epsilon = \psi_r(t-t_0+n^\sigma_{(t_0,t]})$. Hence, from the implication (\ref{eq:impl1}) at $t\ge t_0$ such that $t - t_0 + n^\sigma_{(t_0,t]} = \bar T_{r,\epsilon}$, it follows that
\begin{multline}
  \label{eq:impl2}
  t> t_0 \ge 0, x_0 \in B_r^n, w\in\U\times\S, x\in\T(t_0,x_0,w)\\
  \Rightarrow \quad \alpha(|x(t)|) \le \psi_r(t-t_0+n^\sigma_{(t_0,t]}) + \|w\|
\end{multline}
The proof concludes following exactly the same steps as for the proof of Lemma~2.7 in \cite{sonwan_scl95}.
\bibliographystyle{IEEEtran}
\bibliography{/home/hhaimo/latex/strings.bib,/home/hhaimo/latex/complete_v2.bib,/home/hhaimo/latex/Publications/hernan_v2.bib}
\end{document}